\documentclass[preprint,showpacs,preprintnumbers,twocolumn,amsmath,amssymb,10pt,superscriptaddress]{revtex4-1}
\usepackage{graphicx}
\usepackage{multirow}

\usepackage[breaklinks=true,linkbordercolor={1 1 1}]{hyperref}

\begin{document}

\title{
%% Please do not remove the line below
%\qquad \\ \qquad \\ \qquad \\  \qquad \\  \qquad \\ \qquad \\ 
%% Change title if necessary
KADoNiS-\emph{p}: The astrophysical \emph{p}-process database}

\author{T.~Sz\"ucs}
\email[Corresponding author, electronic address:\\ ]{szucs.tamas@atomki.mta.hu}
\affiliation{Institute for Nuclear Research (MTA Atomki), Debrecen-H-4001, PO Box 51, Hungary}

\author{I.~Dillmann}
\affiliation{GSI Helmholtzzentrum f\"ur Schwerionenforschung GmbH, Darmstadt, Germany}
\affiliation{II. Physikalisches Institut, Justus-Liebig-Universit\"at, Giessen, Germany}

\author{R.~Plag}
\affiliation{GSI Helmholtzzentrum f\"ur Schwerionenforschung GmbH, Darmstadt, Germany}
\affiliation{Institut f\"ur Angewandte Physik, Goethe-Universit\"at, Frankfurt am Main, Germany}

\author{Zs.~F\"ul\"op}
\affiliation{Institute for Nuclear Research (MTA Atomki), Debrecen-H-4001, PO Box 51, Hungary}

\date{September 2013} 

\begin{abstract}
{The KADoNiS-$p$ project is an online database for cross sections relevant to the $p$-process. All existing experimental data was collected and reviewed. With this contribution a user-friendly database using the KADoNiS (\textbf{K}arlsruhe \textbf{A}strophysical \textbf{D}atabase \textbf{o}f \textbf{N}ucleosynthesis \textbf{i}n \textbf{S}tars)  framework is launched, including all available experimental data from (p,$\gamma$), (p,n), (p,$\alpha$), ($\alpha$,$\gamma$), ($\alpha$,n) and ($\alpha$,p) reactions in or close to the respective Gamow window with cut-off date of August 2012 (\url{www.kadonis.org/pprocess}).}
\end{abstract}
\maketitle

\section{INTRODUCTION}

The elements heavier than iron are produced mainly via series of neutron capture reactions and $\beta$-decays. Depending on the neutron density, the path of these processes runs along the valley of stability ($s$-process, \emph{slow} neutron capture) \cite{Kappeler-RMP-2011}, or far away on the neutron rich side of the chart of nuclei involving short lived neutron rich nuclei ($r$-process, \emph{rapid} neutron capture) \cite{Arnould-PR-2007}. However, there are about 32 nuclei on the proton-rich side of the valley of stability which cannot be produced via neutron captures. The main fraction of these so-called "$p$-nuclei" is produced via photodisintegration reactions ($\gamma$-process) on the pre-existing $s$- and $r$-seed nuclei \cite{Woosley-AJS-1978}. First ($\gamma$,n) reactions push the material to the proton-rich side of the valley of stability. At a certain point, ($\gamma$,$\alpha$) and ($\gamma$,p) reactions become dominant diverting the material to lower masses \cite{Rauscher-PRC-2006} where they decay back to stability. Detailed reaction network calculations involving thousands of nuclei and ten thousands of reactions mostly on unstable nuclei still fail to reproduce the observed solar abundance for all $p$-nuclei \cite{Arnould-PR-2003}. The nuclear physics input for these calculations are inferred mainly from Hauser-Feshbach statistical model calculations. To test these predictions systematic measurements in the relevant mass and energy range are needed. Experimentally the inverse radiative capture reactions shall be investigated \cite{Rauscher-PRC-2009}, and from their measured cross sections the reaction rate of the photodisintegration reactions can be calculated applying the principle of detailed balance \cite{Rauscher-IJMPE-2011}. 

The improvement of the predictive power for nuclear reaction cross sections is crucial for further progress in $p$-process models, either by directly replacing theoretical predictions by experimental data or by testing and improving the reliability of statistical models, if the relevant energy range, the so-called Gamow window (GW), is not accessible by experiments. This energy range is typically between $1-6$\,MeV for proton-induced and $3.5-13.5$\,MeV for alpha-induced reactions.

So far, only few reactions have been measured in or close to the relevant energy region. The aim of the KADoNiS-$p$ database is to collect these data, and include them in a user friendly framework for further investigation. An important constraint is also the uniform, comparable datasets with easy access. This database is not only aimed at stellar modellers, but also for experimentalists who want to get a quick overview about the experimental situation from the $p$-process point of view.

\section{THE P-PROCESS DATABASE}
\subsection{History}

The need of a $p$-process database was raised parallel to the first launch of KADoNiS in 2005. In this test phase some (p,$\gamma$), ($\alpha$,$\gamma$), ($\alpha$,n) cross sections on isotopes between $^{56}$Fe and $^{209}$Bi were included. One criteria was to collect only those datasets which contain data points in or close to the GW. There was no extended search for datasets, only some recent data was included.

In July 2010 a new update was launched \cite{Szucs-PoS-2010,Szucs-JPCS-2011}. All experimental data from (p,$\gamma$),  (p,n), ($\alpha$,$\gamma$), and ($\alpha$,n) reactions for targets heavier than $^{70}$Ge had been collected from the EXFOR database \cite{exfor} which have at least one data point below a certain energy. This energy was arbitrarily chosen  to be 1.5 times the upper edge of the respective Gamow window at 3\,GK, which was calculated by the classic equations found in textbooks, e.g. Ref.~\cite{Iliadis-book-2007}.

\subsection{New update}
After the test phase a new update has been prepared, and launched within the framework of KADoNiS. With a cut-off date of August 2012 new datasets from the \mbox{EXFOR} database had been collected, and also the Nuclear Science References (NSR \cite{NSR}) had been reviewed for the most recent data. The criteria was similar in sense of the energy range, except now the relevant energy ranges were taken from a more accurate numerical solution provided by Ref. \cite{Rauscher-PRC-2010}. The list of considered reactions was extended with (p,$\alpha$) and ($\alpha$,p) reactions, too.

\subsection{Web interface}
\begin{figure}[t]
\centering
\includegraphics[width=0.99\columnwidth, clip]{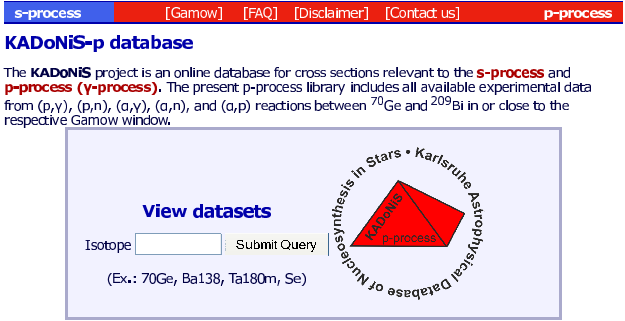}
\caption{Start page the KADoNiS-$p$ database.}
\label{fig:index}
\end{figure}

\begin{figure}[t]
\centering
\includegraphics[width=0.99\columnwidth, clip]{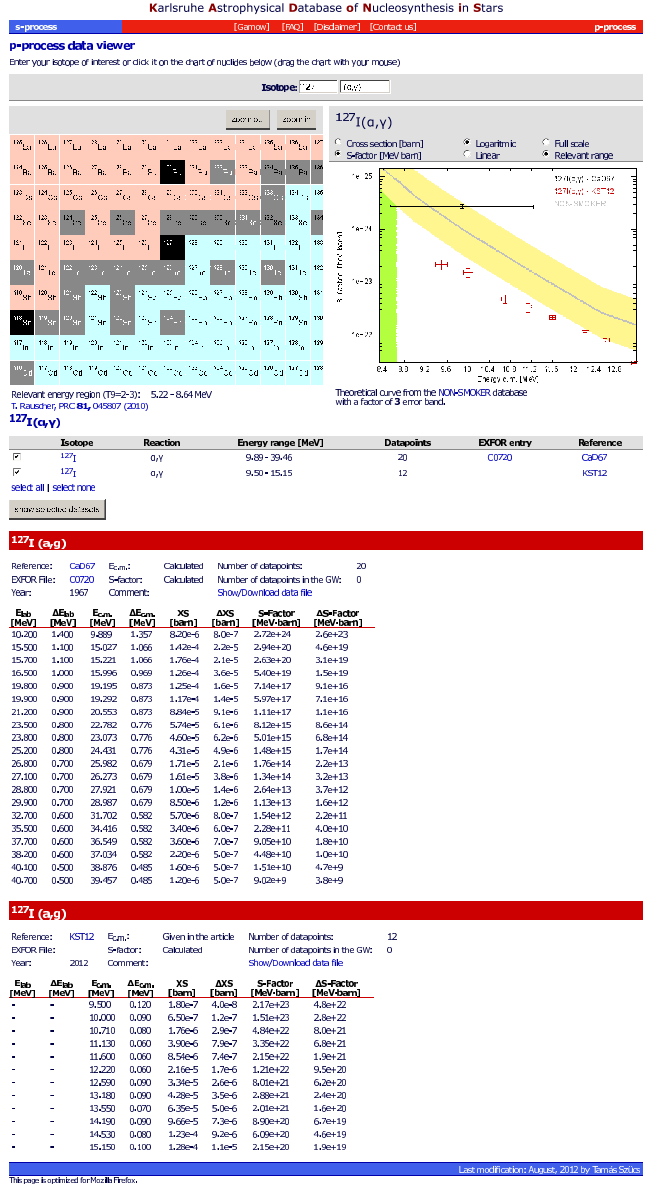}
\caption{Screenshot from the KADoNiS-$p$ database.}
\label{fig:screen}
\end{figure}

The web interface of the $p$-process database has been finalised. Now the whole list of available datasets sorted by reactions can be listed, by pressing return with empty query box (both at the start page fig.~\ref{fig:index} or at the data viewer fig.~\ref{fig:screen}).

The interface is built up as follows: The reaction of interest can be chosen from a list. The isotope can be selected either by entering it directly into the textbox, or from the chart of nuclei. In the chart only those isotopes are highlighted, for which we have a dataset. Below the chart the relevant energy range is shown, and it is also indicated in the plot. The existing datasets are plotted, either the cross section or the astrophysical \mbox{S-factor} can be chosen. Theoretical values for the given reaction from the NON-SMOKER code \cite{non-smoker} are also plotted as a curve, including an uncertainty region of a factor of 2 for proton induced reactions and a factor of 3 for alpha induced reactions.

Below the plotter the list of the existing datasets appears in tabulated form. It shows the energy range of the given dataset, the number of datapoints, the EXFOR reference as a hyperlink (if exists), and also a reference tag formed from the first letters of the names of the authors and the year of the publication. This tag is also a hyperlink, pointing to the source publication where the dataset has been published via its DOI. If this identifier does not exist, then it points to the NSR reference of the article. If neither DOI nor NSR reference exists, then the reference will point to a web page where details about the publication can be found.

Two example dataset can be seen in fig.~\ref{fig:screen}. In the heading the above mentioned identifiers are listed again, and additional information about the E$_{c.m.}$ value and the \mbox{S-factor}. If these quantities are not published in the original article, they are calculated to have uniform datasets. The total number of datapoints and number of datapoints in the Gamow window is also presented. The full dataset is downloadable via a hyperlink.
Finally, the dataset is shown in a tabulated form. In the first two columns the projectile energy in the lab frame (in MeV) and its uncertainty (if available) is given. The 3$^\mathrm{rd}$ and 4$^\mathrm{th}$ columns contain the center-of-mass energy (in MeV). In the next two columns the measured cross section (in barns) is given, and the last two columns show the astrophysical S-factor (in MeV\,barns).

\section{DATA IN THE P-PROCESS DATABASE}
Radiative proton capture datasets are filling the whole mass range. There are more datasets in the lower mass range, where these reactions have higher impact on the predicted $p$-element abundances \cite{Rapp-AJ-2006}.
On the contrary, the radiative $\alpha$ capture reactions have higher impact in the heavier mass range \cite{Rapp-AJ-2006} where only a few experimental data exist up to now.

Neutron emitting reactions are filling almost uniformly the whole mass range. This is due to the fact that in most cases they have much higher cross sections than the radiative capture reactions. Neutron emitting reactions are not entering directly in the reaction network calculations, 
\begin{table}[!h]
\caption{Number of included isotopes and datasets in the KADoNiS-$p$ database sorted by reactions and mass range.}
\label{tab:reactions} 
\centering
\begin{tabular}{l || c c | c c | c c || c c}
\multirow{2}{*}{Reac.} & \multicolumn{2}{c|}{A\,$\leq$\,100} & \multicolumn{2}{c|}{100\,$<$\,A\,$<$\,140} & \multicolumn{2}{c||}{140\,$\leq$\,A} & \multicolumn{2}{c}{{\bf SUM}} \\
 & Isot. & Datas. & Isot. & Datas. & Isot. & Datas. & Isot. & Datas. \\
\hline
(p,$\gamma$)			& 18	& 21	& 15	& 22	& 2	& 3	& {\bf 35}	& {\bf 46} \\
(p,n)						& 25	& 46	& 37	& 88	& 21	& 27	& {\bf 83}	& {\bf 161} \\
($\alpha$,$\gamma$)	& 5	& 5	& 9	& 13	& 4	& 5	& {\bf 18}	& {\bf 23} \\
($\alpha$,n)			& 11	& 24	& 16	& 22	& 12	& 26	& {\bf 39}	& {\bf 72} \\
($\alpha$,p)			& 3	& 3	& 3	& 4	& --	& --	& {\bf 6}	& {\bf 7} \\
\end{tabular}
\end{table}
but they help to finetune the statistical model calculations since those have to describe both the radiative capture and the neutron emitting reactions with one parameter set.

No (p,$\alpha$) dataset matched the criteria of the energy range so far, and from the inverse ($\alpha$,p) reaction datasets exist only in the lower mass region, where they have a smaller impact.

\section{FURTHER PLANS}

Implementation of further theoretical data is on the way (TALYS \cite{talys}). In addition, the database will be extended with partial cross sections from reactions leading to metastable states. Based on the existing datasets 7-parameter reaction rate fits will be derived, which can be included in reaction libraries.
\\

\section{CONCLUSIONS}
The new user-friendly p-process database is launched, it is the first of its kind. This database is aimed to give a quick overview about the experimental situation for the \mbox{$p$-process studies}. All the included datasets are relevant to the mass, energy and temperature of the $p$-process.

\subsection*{Acknowledgement}
The authors are indebted to Dr. Thomas Rauscher for giving the permission of including the data directly from the NON-SMOKER database.
This work was supported by the OTKA K101328, NN83261 \mbox{(EuroGENESIS)} projects, and by the Helmholtz Association via the Young Investigators projects VH-NG-627 and VH-NG-327.

\end{document}